\documentclass[twocolumn,showpacs,amsmath,amssymb,pre]{revtex4}
\usepackage{bbm}

\usepackage{graphicx}%
\usepackage{dcolumn}
\usepackage{amsmath}
\usepackage{bm}
\usepackage{color}
\usepackage{subeqnarray}
\newcommand{\be}{\begin{equation}}
\newcommand{\ee}{\end{equation}}
\newcommand{\bey}{\begin{eqnarray}}
\newcommand{\eey}{\end{eqnarray}}
\newcommand{\bw}{\begin{widetext}}
\newcommand{\ew}{\end{widetext}}

\newcommand{\ra}{\rangle}
\newcommand{\la}{\langle}

\newcommand{\ba}{\begin{array}}
\newcommand{\ea}{\end{array}}
\newcommand{\bi}{\begin{itemize}}
\newcommand{\ei}{\end{itemize}}
\newcommand{\bem}{\begin{enumerate}}
\newcommand{\eem}{\end{enumerate}}

\begin{document}

\title {Complexity and instability of quantum motion near a quantum phase
transition}

\author{Pinquan Qin}
\affiliation{Department of Modern Physics, University of Science and Technology
of China, Hefei 230026, China}
\author{Wen-ge Wang} \email{wgwang@ustc.edu.cn}
\affiliation{Department of Modern Physics, University of Science and Technology of China,
Hefei 230026, China}
\author{Giuliano Benenti}
\affiliation{CNISM and Center for Nonlinear and Complex Systems,
Universit\`a degli Studi dell'Insubria, Via Valleggio 11, 22100 Como, Italy}
\affiliation{Istituto Nazionale di Fisica Nucleare, Sezione di Milano,
via Celoria 16, 20133 Milano, Italy}
\author{Giulio Casati}
\affiliation{CNISM and Center for Nonlinear and Complex Systems,
Universit\`a degli Studi dell'Insubria, Via Valleggio 11, 22100 Como, Italy}
\affiliation{Istituto Nazionale di Fisica Nucleare, Sezione di Milano,
via Celoria 16, 20133 Milano, Italy}

 \date{\today}

 \begin{abstract}
We show that the number of harmonics of the Wigner function,
recently proposed as a measure of quantum complexity, can also be used to
characterize quantum phase transitions. The non-analytic
behavior of this quantity in the neighborhood of
a quantum phase transition is illustrated by means of the Dicke model
and is compared to two well-known measures of the (in)stability of quantum motion:
the quantum Loschmidt echo and fidelity.
 \end{abstract}


 \maketitle

 \section{Introduction}

Quantifying the complexity of quantum systems
is a challenging fundamental problem, also of practical relevance
to a better understanding of the minimum computational resources
required to simulate many-body quantum systems.
A very convenient framework for investigating quantum complexity is the
phase-space representation of quantum
mechanics~\cite{Chirikov81,Gu90,Ford91,Brumer97,Gong03,Sokolov08,Benenti09,SZ08,Vinitha10,Casati12,Benenti12}.
The main advantage of such an approach is that it can be equally well applied to
classical and quantum mechanics, being based on the structure
of the Liouville density in the former case and of the Wigner function
in the latter. In this context, the number of Fourier harmonics
of the Wigner function ---
whose growth rate reproduces in the semiclassical limit the well-known
notion of classical complexity based on the local exponential
instability of chaotic dynamics --- has already been used to measure
the complexity of single-particle~\cite{Sokolov08,Benenti09,SZ08} and
many-body~\cite{Vinitha10} quantum systems.

 An interesting question is
whether the number of harmonics is capable of characterizing the
 complexity of the quantum motion near quantum phase transitions (QPTs),
 where a slight variation in the controlling parameter may induce dramatic change in the wave function.
 As is known, in addition to conventional quantities such as the order parameter,
 the fidelity, which is the most basic metric quantity, is also capable of characterizing QPTs
 \cite{zan2006,zan2007,zana2007,coz2007,cozz2007,buo2007,ven2007,zho2008,liu2009,gu2010,rams11}.
In the context of QPTs, fidelity is defined as the overlap of the ground states of the
 Hamiltonians with a slight difference in the controlling parameter driving the QPT.
 The dramatic change of the wave function at a QPT implies a fast decrease of the fidelity
 when approaching the critical point.
 Another metric quantity useful for characterizing the occurrence of a QPT is the
quantum Loschmidt echo (LE)~\cite{peres84,nie2000,ben2004,jal2001,jac2001,
cer2002,jac2002,cuc2002,STB03,wan2005,pro2002t,pro2002,BC02,VH03,wan2004,wang2005,
Gorin-rep,wan2007,jacquodreport,Wang10}, which provides a measure to
the stability of the quantum motion under two slightly different Hamiltonians.
This quantity exhibits a dramatic change in its decaying behavior in the
neighborhood of critical points.

 Since the number of harmonics is geometric in nature, being
based on the richness of the phase-space structure, it is
natural to expect that it could be related to other metric quantities such as  the
fidelity and the LE and, as a result, could also be used to characterize QPTs.
 It should be stressed that, in contrast to the number of
harmonics, the fidelity and the LE in general are not good measures
of the complexity of quantum motion.
For example, the LE  decay does not clearly distinguish,
either in quantum or in classical mechanics, between
chaotic and integrable systems and for integrable
systems the decay can be even faster than for chaotic
systems in certain situations~\cite{Gorin-rep}.
On the other hand, in classical mechanics the number of harmonics
of the classical phase-space distribution function grows linearly
for integrable systems and exponentially for chaotic systems, with
the growth rate related to the rate of local exponential instability
of classical motion~\cite{Brumer97,Gong03}. Thus the growth rate
 of the number of harmonics is a measure of classical complexity,
whose extension to the quantum realm has been more recently
demonstrated in Refs.~\cite{Sokolov08,Benenti09,SZ08,Vinitha10}.

In this paper we show that the number of harmonics of the Wigner function
is useful also in characterizing QPTs.
 For this purpose we study a Hamiltonian that in a low-energy region relevant to a QPT
 can be written as a finite number of harmonic oscillators,
 with at least one mode whose frequency vanishes at the critical point.
 In particular, we focus on the Dicke model in the
 thermodynamic limit, with a single zero mode.
 We show that the number of harmonics diverges when approaching the QPT,
 which is a manifestation of the fact that a small
variation of the controlling parameter provides a dramatic
change of the wave function in the vicinity of the critical point.
As a consequence, the number of harmonics
exhibits a nonanalytic behavior at the QTP,
similarly to other metric quantities such as the fidelity and the LE,
which we briefly discuss for the Dicke model.

The paper is organized as follows. In Sec.~\ref{sec:basics}
we review basic definitions of the harmonics of the Wigner
function and discuss their properties for a Hamiltonian describing a QPT
in terms of a finite number of harmonic oscillators.
The time evolution of the number of harmonics is then
illustrated for the Dicke model in the vicinity of its QPT
in Sec.~\ref{sec:dickeharmonics}, while other metric quantities,
i.e., the LE and the fidelity, are discussed
in Sec.~\ref{sec:dickefidelity}.
We conclude with a summary in Sec.~\ref{sec:conc}.

\section{Harmonics of the Wigner function close to a QPT}
\label{sec:basics}

In this section we recall the definition of the number of harmonics of the
Wigner function and discuss its application in the neighborhood of a QPT.
For the sake of simplicity, we limit ourselves to
systems whose Hamiltonian can be written in terms of a set
of bosonic creation-annihilation operators, that is to say,
 \be
 \hat{H}\equiv \hat{H}^{(0)}+\hat{H}^{(1)}, \label{H01}
 \ee
 where $\hat{H}^{(0)}=\hat{H}^{(0)}(\hat{n}_{1},\ldots,\hat{n}_{M})$
is a time-independent unperturbed Hamiltonian and
 $\hat{H}^{(1)}=\hat{H}^{(1)}({\hat{a}_{1}^{\dag},
\ldots,\hat{a}_{M}^{\dag}},\hat{a}_{1},\ldots,
 \hat{a}_{M};t)$ is a perturbation.
Here $\hat{a}_{i}^{\dag}$ and $\hat{a}_{i}$ are bosonic creation
and annihilation operators, $\hat{n}_{i}=\hat{a}_{i}^{\dag}\hat{a}_{i}$
are particle number operators,
and $M$ is the number of bosonic modes.
Using the $c$-number $\bm{\alpha}$-phase
method~\cite{Bargmann,Glauber,Agarwal},
the Wigner function of a state, which is described
 by a density operator $\hat{\rho}(t)$, can be written as
 \begin{gather}
 W(\bm{\alpha},\bm{\alpha}^{\ast};t)=
\frac{1}{\pi^{2M}\hbar^{M}}\int d^{2}\bm{\chi}
 {\rm exp}\Big(\frac{\bm{\chi^{\ast}}\cdot\bm{\alpha}}{\sqrt{\hbar}}
- \frac{\bm{\chi} \cdot
 \bm{\alpha}^{\ast}}{\sqrt{\hbar}}\Big) \notag \\
 \times {\rm Tr}[\hat{\rho}(t)\hat{D}(\bm{\chi})],
 \label{w}
 \end{gather}
 where $\bm{\alpha}$ and $\bm{\chi}$ are $M$-dimensional
complex parameter vectors and
 \be
 \hat{D}(\bm{\chi}) ={\rm exp}\Big[ \sum_{i=1}^{M} \big( {\chi_{i}
 \hat{a}_{i}^{\dag} - \chi_{i}^{*} \hat{a}_{i}} \big) \Big]
 \ee
 is the so-called displacement operator.
 Coherent states $|\bm{\alpha}\rangle$
can be generated by $\hat{D}$:
 \be
 |\bm{\alpha}\rangle \equiv |\alpha_{1} \alpha_{2}
 \alpha_{3}\cdots\alpha_{M}\rangle = \hat{D}\Big(
 \frac{\bm{\alpha}}{\sqrt{\hbar}}\Big) |00\cdots0\rangle,
 \ee
with $|\alpha_i\rangle$ being the eigenstate of the annihilation
operator $\hat{a}_i$, namely,
$\hat{a}_i|\alpha_i\rangle=(\alpha_i/\sqrt{\hbar}) |\alpha_i\rangle$,
and ${|00\cdots0\rangle}$ being the vacuum state.

We then consider
 the amplitudes, denoted by $W_{\bm{m}}(\bm{I};t)$,
of the $M$-dimensional Fourier expansion of the Wigner
 function:
 \be
 W(\bm{\alpha},\bm{\alpha}^{\ast};t)=\frac{1}{\pi^{M}}\sum_{\bm{m}}W_{\bm{m}}(\bm{I};t)
 e^{i\bm{m}\cdot\bm{\theta}},
 \ee
 where $\bm{m}$ is an $M$-dimensional integer vector.
 Here $\bm{I}$ and $\bm{\theta}$ are $M$-dimensional real vectors,
 determined from $\bm{\alpha}$ by the relation
 $\alpha_{l}=\sqrt{I_{l}}e^{-i\theta_{l}}$, with $l=1,\ldots,M$.

To estimate the number of harmonics~\cite{Sokolov08,Benenti09}, we
will consider
$\sqrt{\langle\bm{m^{2}}\rangle_{t}}$, with
$\langle\bm{m^{2}}\rangle_{t}$ the second moment of
the harmonics distribution~\cite{footnote_entropy},
 \be
\langle\bm{m^{2}}\rangle_{t}=\sum_{\bm{m}} |\bm{m}|^2
\mathcal{W}_{\bm{m}}(t) , \ee
 where
 \be
 \mathcal{W}_{\bm{m}}(t)\equiv\frac{\int d\bm{I}
|W_{\bm{m}}(\bm{I};t)|^{2}}{\sum_{\bm{m}}
 \int d\bm{I}|W_{\bm{m}}(\bm{I};t)|^{2}}.
 \ee
 It is useful to give an explicit expression of $\mathcal{W}_{\bm{m}}(t)$
in terms of the density matrix $\hat{\rho}$.
 For this purpose, one may note that in the basis of $\hat{H}^{(0)}$,
 namely, in $|\bm{n}\rangle=|n_{1}\cdots n_{M}\rangle$,
 the displacement operator $\hat{D}(\bm{\chi})$ has the well-known
matrix elements~\cite{schwinger53}
 \be
 \langle n_{i}+m_{i}|\hat{D}(\chi_{i})|n_{i}\rangle =
\sqrt{\frac{n_{i}!}{(n_{i}+m_{i})!}}
 \chi_{i}^{m_{i}}e^{-|\chi_{i}|^{2}/2}
 L_{n_{i}}^{m_{i}}(|\chi_{i}|^{2})
 \ee
 $(n_{i},m_{i}\geqslant0,i=1,\ldots,M)$, where $L_{n_{i}}^{m_{i}}(x)$ indicate
 Laguerre polynomials.
 Using this expression, the integration over $\bm{\chi}$
in Eq.~(\ref{w}) can be carried out.
 Then, making use of the orthogonality and the completeness of
 Laguerre polynomials, one has
 \bey
 W_{m_{i}}(I_{i};t) &=& \frac{2}{\hbar} e^{-\left(2/\hbar\right)
I_{i}}
 \sum_{n_{i}=0}^{\infty} (-1)^{n_{i}}
 \sqrt{\frac{n_{i}!}{(n_{i}+m_{i})!}}
\left(\frac{4I_{i}}{\hbar}\right)^{m_{i}/2}\nonumber \\
 &&\times L_{n_{i}}^{m_{i}}\left(\frac{4I_{i}}{\hbar}\right) \langle
 n_{i}+m_{i}|\hat{\rho}(t)|n_{i}\rangle.
 \eey
 This gives the following expression of
$\mathcal{W}_{\bm{m}}(t)$~\cite{Vinitha10}:
 \be
 \mathcal{W}_{\bm{m}}(t)=
\frac{\sum_{\bm{n}}|\langle\bm{n+m}|\hat{\rho}(t)|\bm{n}\rangle|^{2}}
 {\sum_{m_{1},\cdots,m_{M}\geq0}
\sum_{\bm{n}}|\langle\bm{n+m}|\hat{\rho}(t)|\bm{n}\rangle|^{2}}.
 \label{wm}
 \ee

In fact, as only the lowest-energy levels are concerned close to the critical point,
we assume that the Hamiltonian describing a QPT can be approximately written
in terms of $M$ harmonic oscillators (up to an irrelevant constant
energy term)
\begin{equation}\label{Hhar}
\hat{H}(\lambda)
=\sum_{i=1}^M e_{i}(\lambda)
\hat{c}_{i}^{\dag}(\lambda) \hat{c}_{i}(\lambda),
\end{equation}
where $\lambda$ is the controlling parameter driving the QPT and
$\hat{c}_i^{\dag}(\lambda)$ and $\hat{c}_i(\lambda)$ are
bosonic creation and annihilation operators for the $i$th mode,
with frequency $e_i(\lambda)/\hbar$.
We use $|\Phi_{\bm{\mu}}(\lambda)\ra$, with $\bm{\mu}=(\mu_1,\ldots,\mu_M)$
and {$\mu_1,\ldots,\mu_M=0,1,\ldots$},
to denote eigenstates of $\hat{H}(\lambda)$:
 \be \hat{H}(\lambda) |\Phi_{\bm{\mu}}(\lambda)\ra =
E_{\bm{\mu}}(\lambda) |\Phi_{\bm{\mu}}(\lambda)\ra,
\label{bz} \ee
where
 \bey \label{Emu}
 |\Phi_{\bm{\mu}}(\lambda)\ra &=&
\prod_{i=1}^M\frac{1}{\sqrt{\mu_i!}} (\hat{c_i}^\dag)^{\mu_i} |\Phi_{0}
 (\lambda)\ra,
 \\ E_{\bm{\mu}}(\lambda) &=& \sum_i\mu_i e_i({\lambda}),
 \eey
 with $|\Phi_0(\lambda)\rangle$ indicating the ground state
of the Hamiltonian (\ref{Hhar}).

Let us consider two values $\lambda_0$ and $\lambda$ of the
controlling parameter.
For the sake of clarity, while using $\bm{\mu}$ to indicate eigenstates
$|\Phi_{\bm{\mu}}(\lambda)\ra $ of $\hat{H}(\lambda)$ we use
$\bm{n}=(n_1,\ldots,n_M)$ to label the eigenstates
$|\Phi_{\bm{n}}(\lambda_0)\ra $ of $\hat{H}(\lambda_{0})$.
We consider the time evolution
driven by the Hamiltonian $\hat{H}(\lambda)$,
starting at a time $t=0$ from the ground state
$|\Phi_0(\lambda_0)\rangle$ of $\hat{H}(\lambda_{0})$.
Substituting the time-dependent state vector
 \be
 |\Psi(t)\rangle = \sum_{\bm{\mu}}
e^{-iE_{\bm{\mu}}(\lambda)t} |\Phi_{\bm{\mu}}(\lambda)\rangle\langle
 \Phi_{\bm{\mu}}(\lambda) |\Phi_{0}(\lambda_{0})\rangle
 \label{Psit} \ee
 into Eq.~(\ref{wm}),
the second moment of the harmonics
distribution can be expressed as follows:
 \be
 \langle \bm{m}^{2}\rangle_{t}=\frac{\sum\limits_{\mbox{\tiny
 $\begin{array}{c}m_1,\ldots,m_M\geq0\end{array}$}}
 \bm{m}^{2} \mathcal{F}(\bm{m},t)}
 {\sum\limits_{\mbox{\tiny
 $\begin{array}{c}m_1,\ldots,m_M\geq0\end{array}$}}
 \mathcal{F}(\bm{m},t)},
 \label{wt}
 \ee
 where
 \be
 \mathcal{F}(\bm{m},t)=\sum_{n_1,\ldots,n_M\geq0}
\Big|\sum\limits_{\bm{\mu}\bm{\nu}} K_{\bm{\mu\nu}}^{\bm{nm}}
 e^{-i(E_{\bm{\mu}}-E_{\bm{\nu}})t} \Big|^{2},
 \label{f}
 \ee
 \be
\begin{array}{c}
K_{\bm{\mu\nu}}^{\bm{nm}} = \langle\Phi_{\bm{n+m}}(\lambda_{0})
 |\Phi_{\bm{\mu}}(\lambda)\rangle
\langle\Phi_{\bm{\mu}}(\lambda)| \Phi_{0}(\lambda_{0})\rangle
\\\\
 \times \langle \Phi_{0}(\lambda_{0})|
 \Phi_{\bm{\nu}}(\lambda)\rangle \langle\Phi_{\bm{\nu}}(\lambda)|
 \Phi_{\bm{n}}(\lambda_{0})\rangle.
\end{array}
 \label{Q}
 \ee
 In computing the harmonics, we have taken $\hat{H}^{(0)}=\hat{H}(\lambda_0)$ and
$\hat{H}=\hat{H}(\lambda)$.

Since the second moment of the harmonics distribution is written in terms
of inner products of eigenstates at different values of the controlling
parameter, we expect this quantity to change dramatically with
$\lambda$ and $\lambda_0$ approaching the critical point $\lambda_c$.
That is, due the fast change of the wave function at a QPT, we expect
the number of harmonics to be able to detect the QPT.
In the following section we shall illustrate such a property
in the physically relevant example of the Dicke model.

\section{The Dicke model QPT}
\label{sec:dickeharmonics}

\subsection{Dicke Hamiltonian in the thermodynamic limit}

 The single-mode Dicke model~\cite{Emary03}
 describes the interaction between a single bosonic mode and a collection of $N$ two-level atoms.
 The system can be described in terms of the collective operator ${\bf \hat{J}}$ for the $N$
 atoms, with
 \be
 \hat{J}_{z} \equiv \sum_{i=1}^{N} \hat{s}_{z}^{(i)};\ \ \hat{J}_{\pm} \equiv \sum_{i=1}^{N}
 \hat{s}_{\pm}^{(i)},
 \ee
 where $\hat{s}_{x (y,z)}^{(i)}$ are Pauli matrices divided by $2$ for the $i$th atom.
 The Dicke Hamiltonian is written as~\cite{Emary03} (hereafter
we take $\hbar =1$)
 \be
 \hat{H}(\lambda )=\omega_{0}\hat{J}_{z}+\omega \hat{a}^{\dag}\hat{a} + ({\lambda}/{\sqrt{N}})
 (\hat{a}^{\dag}+\hat{a})(\hat{J}_{+}+\hat{J}_{-}).
 \label{DH} \ee
 Performing the Holstein-Primakoff transformation
 \begin{gather}
 \hat{J}_{+} = \hat{b}^{\dag}\sqrt{2j - \hat{b}^{\dag}\hat{b}},\ \  \hat{J}_{-} =
 \Big(\sqrt{2j-\hat{b}^{\dag}\hat{b}}\Big)\hat{b}, \notag \\
 \hat{J}_{z} = (\hat{b}^{\dag}\hat{b}-j),
 \end{gather}
 one has
 \bey
 \hat{H}(\lambda) &=&\omega_{0}(\hat{b}^{\dag}\hat{b}-j)+\omega \hat{a}^{\dag}\hat{a} \nonumber \\
 &+&\lambda(\hat{a}^{\dag}+\hat{a})\left(\hat{b}^{\dag}\sqrt{1-\frac{\hat{b}^{\dag}\hat{b}}{2j}}
 +\sqrt{1-\frac{\hat{b}^{\dag}\hat{b}}{2j}}\ \hat{b}\right), \ \
 \label{H}
 \eey
 where $j={N}/{2}$.
 As shown in Ref.~\cite{Emary03},
 in the thermodynamic limit $N\to \infty$, the system undergoes a QPT at
 $\lambda_c = \frac 12 \sqrt{\omega \omega_0}$, with a normal phase for $\lambda <\lambda_c$
 and a superradiant phase for $\lambda > \lambda_c$.

 In the normal phase and for low-lying states, $\hat{b}^{\dag}\hat{b}/N$
can be neglected
 and then the Hamiltonian in Eq.~(\ref{H}) can be written as
 \bey
 {\hat{H}(\lambda)} &=&\omega_{0}\hat{b}^{\dag}\hat{b}+ \omega \hat{a}^{\dag}\hat{a} \nonumber \\
 &&+ \lambda (\hat{a}^{\dag}+ \hat{a})(\hat{b}^{\dag}+\hat{b})- j\omega_{0}.
 \label{h1}
 \eey
 This Hamiltonian can be diagonalized and written as a sum of two
harmonic oscillators
 \bey
 {\hat{H}(\lambda)} &=& e_{1}({\lambda}) \hat{c}_{1}^{\dag}\hat{c}_{1} +
 e_{2}(\lambda) \hat{c}_{2}^{\dag}\hat{c}_{2} + g,
 \label{Hs}
 \eey
 where $g$ is a $c$-number function and
 \be
 e_{1,2}({\lambda}) =
\left\{\frac{1}{2}\left[(\omega^{2} + \omega^{2}_{0})
\pm \sqrt{(\omega^{2}_{0}-\omega^{2})^{2}
  + 16\lambda^{2}\omega\omega_{0}}\right]\right\}^{1/2},
  \label{ee}
 \ee
 with $e_1({\lambda}) < e_2({\lambda})$.
 It is seen that $e_{1}({\lambda})=0$
and $e_{2}({\lambda})\ne 0$ for $\lambda= \lambda_c$,
 hence the ground level of $\hat{H}(\lambda_c)$ is infinitely degenerate
 and the system undergoes a QPT at $\lambda_c$, with a single zero mode.

 In the superradiant phase,
 one may write the bosonic modes in Eq.~(\ref{H}) in terms of two
 new bosonic modes
 \be
{ \hat{a}^\dag\rightarrow \hat{a}^{\prime\dag} + \sqrt{A},\
\ \ \hat{b}^\dag\rightarrow \hat{b}^{\prime\dag} - \sqrt{B},}
 \label{ab}
 \ee
 where  {$A$ and $B$} are of order $N$.
 Then, taking the thermodynamic limit and eliminating the
linear terms in the Hamiltonian by choosing
 appropriate values of  {$A$ and $B$}, one can also
diagonalize the Hamiltonian in the superradiant phase,
resulting in the same form as in Eq.~(\ref{Hs}),
 with the following energy levels of two modes:
 \be
e_{1,2}(\lambda) = \left\{\frac{1}{2}\left[\omega^{2} +
 {\frac{\omega^{2}_{0}}{\kappa^2}} \pm
\sqrt{\left( {\frac{\omega^{2}_{0}}{\kappa^2}}-\omega^{2}\right)^{2}
  + 4\omega^2\omega^2_{0}}\right]\right\}^{1/2},
 \label{ep}
 \ee
 where $ {\kappa\equiv \omega \omega_0/4\lambda^2}$ and
$e_{1}(\lambda)<e_{2}(\lambda)$.
 It is easy to see that $e_{1}(\lambda)=0$ and
$e_{2}({\lambda})\ne 0$ for $\lambda= \lambda_c$,
hence the ground level of $\hat{H}(\lambda_c)$ is also infinitely
degenerate (and with a single zero mode)
at the critical point from the  {superradiant phase} side.

Therefore, the Dicke Hamiltonian close to the critical point
can be approximated, in both phases, by an effective Hamiltonian
of the form (\ref{Hhar}) with $M=2$ harmonic oscillators.
If both values $\lambda$ and $\lambda_0$
of the controlling parameter belong to the same phase, as we will
always consider in our paper, then we can compute the
second moment of the harmonic distribution by means of
Eqs.~(\ref{wt})-(\ref{Q}).

\subsection{Time evolution of the number of harmonics of the Wigner
function near the critical point}
 \label{sect-period-nh}

When $\lambda$ is sufficiently close to the critical point
$\lambda_c$ and the dynamics is mainly determined by the
lowest-energy levels, i.e., those of the zero mode, Eq.~(\ref{Hhar})
can be further simplified to a single harmonic oscillator ($M=1$).
We have numerically checked that adding the second mode, i.e., the nonzero one,
does not significantly affect the dynamics close to the critical point.
Therefore, in what follows we will present data for the single mode only.
We would like to point out that, even though the Dicke model
in the vicinity of the QPT
has the simple form of a one-dimensional (1D) harmonic oscillator,
this does not imply triviality of the problem considered here because
the oscillator has different frequencies at different values
of $\lambda$ and in particular it is completely degenerate at the
critical point $\lambda_c$.

\begin{figure}
 \includegraphics[width=\columnwidth]{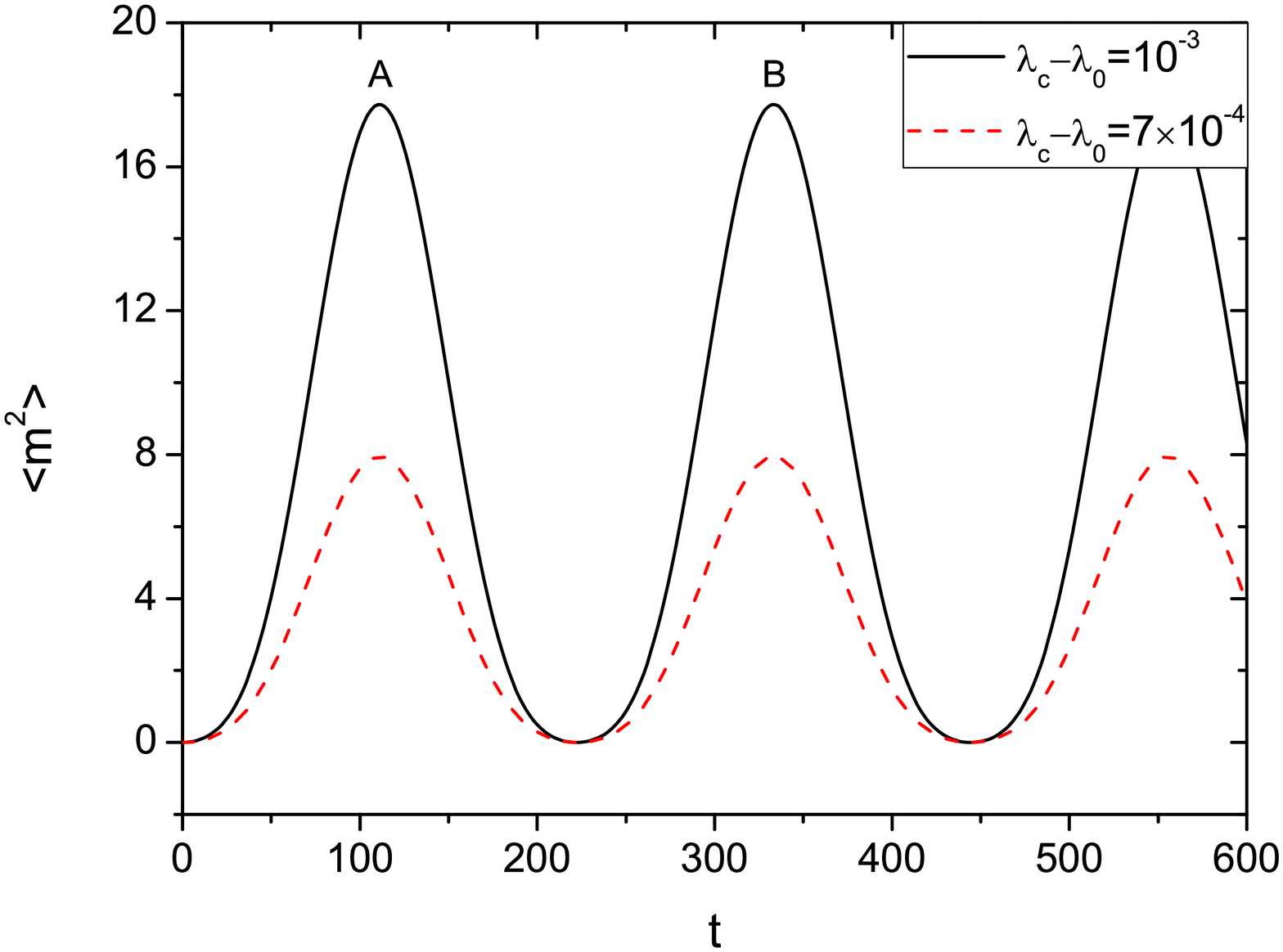}
 \caption{(Color online) Second moment $\langle m^{2}\rangle_{t}$
of the harmonic distribution
 versus time $t$ for the Dicke model at
$\lambda=\lambda_{c}-10^{-4}$,
$\lambda_{0}=\lambda_{c}-10^{-3}$ (solid curve) and
$\lambda_0=\lambda_{c} - 7\times 10^{-4}$ (dashed curve).
Hereafter we set in numerical calculations $\omega=\omega_0=1$.}
 \label{wqhcdl}
 \end{figure}

For this model, as shown in Fig.~\ref{wqhcdl}, the basic feature of the
second moment $\langle m^2 \rangle_t$
of the harmonics distribution of the Wigner function
is that it is a periodic function of the time $t$.
This is because, as discussed above, the Hamiltonian $\hat{H}(\lambda)$ has
effectively the form of a 1D harmonic oscillator.
To find the period, let us consider the times
corresponding to the maximum values of $\la m^2\ra_t$,
e.g., points $A$ and $B$ in Fig.~\ref{wqhcdl},
and denote these times by $t_p$.
These times $t_p$ can be found by the requirement
$ d\langle m^{2} \rangle_{t} /dt= 0$.
Making use of Eq.~(\ref{wt}), we find that
$d\langle m^{2} \rangle_{t} /dt= 0$ is equivalent to
the relation
 \be
\begin{array}{c}
 \sum_{\vartheta\vartheta'} G
 \sin [(\mu-\nu-\beta+\gamma)e_1(\lambda) t] \
\\ \\
\times \cos [(\mu'-\nu'-\beta'+\gamma')
 e_1(\lambda)t] =0,  \label{sumG}
\end{array}
\ee
where ${\vartheta} \equiv ({\mu,\nu,\beta,\gamma})$
and $G$ is a function of the quantity
$K_{\mu\nu}^{nm}$ defined in Eq.~(\ref{Q}).
As discussed in Appendix A,
all of $\mu,\ \nu,\ \beta$, and $\gamma$ are even numbers,
hence the left-hand side of Eq.~(\ref{sumG}) is zero at the
times satisfying
$e_1(\lambda)t=k\pi/2$, with odd numbers
$k$ corresponding to the maximums and
even numbers $k$ corresponding to the
minimums of $\la m^2\ra_t$.
Hence, we have the maximums at the times
 \be
 t_{p}=\frac{k\pi}{2e_1(\lambda)}\ \ \ (k=1,3,5,\ldots).
 \label{tp}
 \ee
Therefore, $\langle m^2 \rangle_t$ has a period
 \be T=\frac{\pi}{e_{1}({\lambda})}. \label{T-nh} \ee
 This value is in agreement with the numerical calculations shown
in Fig.~\ref{wqhcdl}.
Note that this period $T$ is half of the period of the lowest mode of the system
$\hat{H}(\lambda)$ and is infinitely long at the critical point $\lambda_c$.
It has the same form in both phases of the Dicke model.
It is interesting to remark that the period $T$ of the number
of harmonics can be related to the gap between the ground state and
the first excited state via Eq.~(\ref{tp}). Such an important quantity
can be obtained also by other time-dependent metric quantities
such as the quantum Loschmidt echo (see Sec.~\ref{sec:dickefidelity}
below).

\begin{figure}
 \includegraphics[width=\columnwidth]{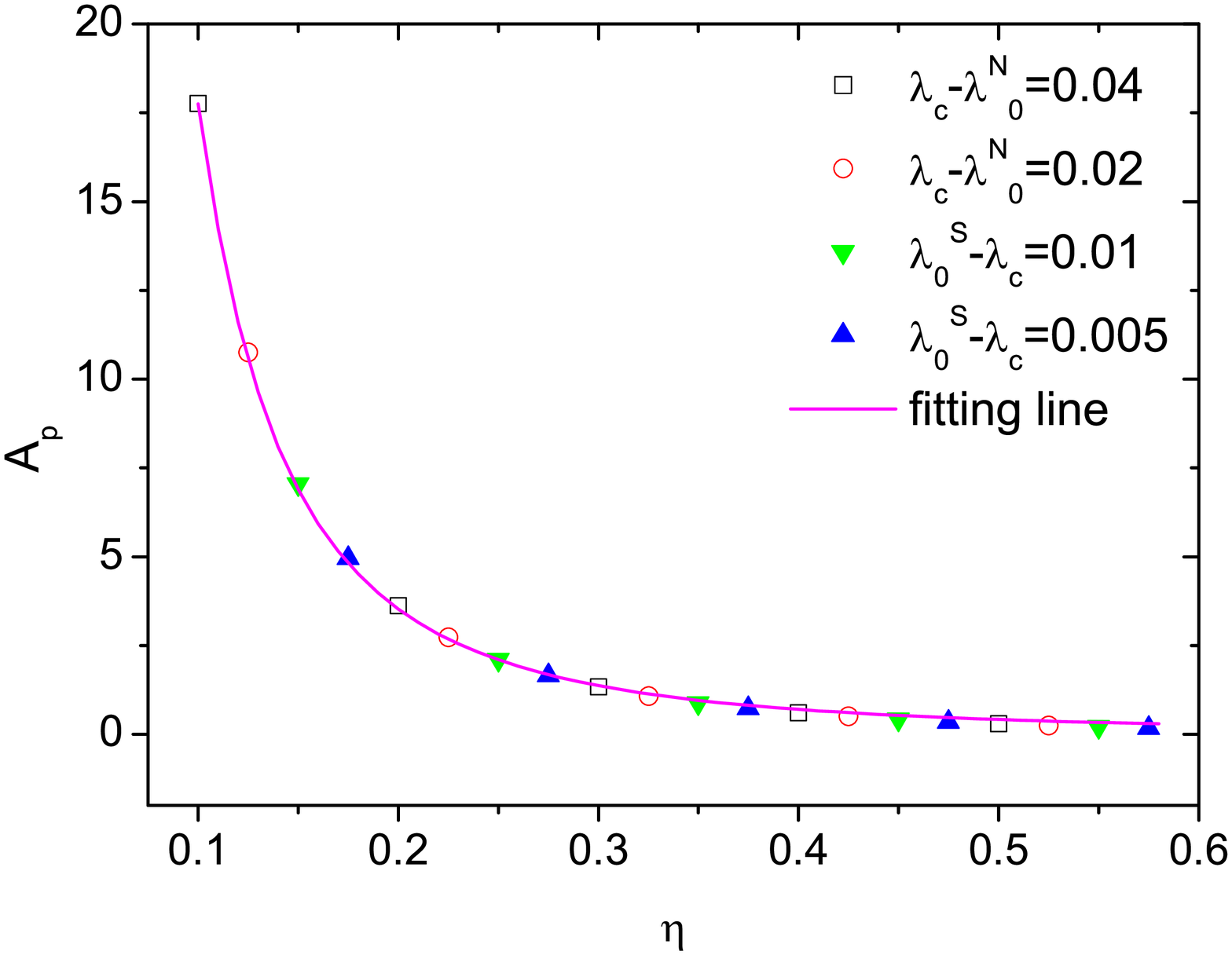}
 \caption{(Color online) Dependence of the amplitude $A_p$ on $\eta$ for
 different values of $|\lambda_c-\lambda_0|$ and $|\lambda_c-\lambda|$.
Open symbols (with the superscript $N$ on $\lambda_0$)
stand for the normal phase and closed symbols (superscript $S$) for the
superradiant phase.
The solid fitting curve is
given by $A_p=a|\eta|^b$, with $a=0.083$ and $b=-2.3$.}
 \label{apt}
 \end{figure}

Next we consider the amplitude of the oscillations of
the number of harmonics. We study the maximum value  of $\la m^2\ra_t$ in
Eq.~(\ref{wt}), which is determined by
the functions $\mathcal{F}(m,t_{p})$.
Substituting Eq.~(\ref{tp}) into Eq.~(\ref{f}), we obtain
 \be \mathcal{F}(m,t_{p})=\sum\limits_{n\geq0}\Big|\sum\limits_{\mu\nu} (-1)^{(\mu-\nu)/2}
K_{\mu\nu}^{nm} \Big|^{2}. \ee
Therefore, the amplitude of $\langle m^{2} \rangle_{t}$, denoted by $A_p$,
is obtained from the summation of the terms $K_{\mu\nu}^{nm}$,
with weights $(-1)^{(\mu-\nu)/2}$.
As discussed in Appendix A,
the quantities $K_{\mu\nu}^{nm}$ are
in both phases functions of the ratio
\be
\eta=\frac{\lambda-\lambda_c}{\lambda_0-\lambda_c}.
\label{eq:etascaling}
\ee
Therefore, $A_p$ depends only on the ratio $\eta$ and not
on $\lambda$ and $\lambda_0$ separately.
This property
(illustrated in Fig.~\ref{apt}) is in agreement with the
scaling derived in Ref.~\cite{QPTscaling} for metric
quantities for QPTs with a single bosonic zero
mode at the critical point.
As shown in  Fig.~\ref{apt}, the dependence of $A_p$ on $\eta$ can be
well fitted by the curve $A_p =a |\eta|^b$ and
since $b$ is negative, $A_p$ diverges when $|\eta|$ goes to zero.

\begin{figure}
 \includegraphics[width=\columnwidth]{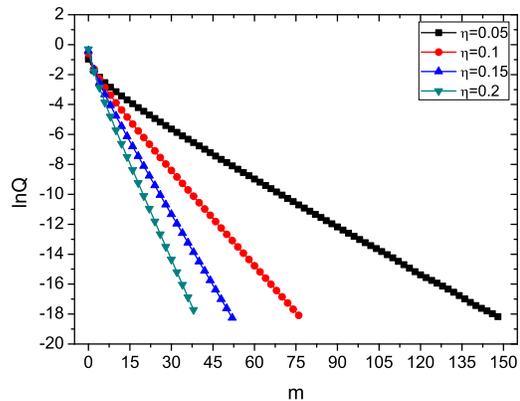}
 \caption{(Color online) Variation of $\ln Q$ with $m$
for $\lambda_c-\lambda=0.01$ and the harmonics
distribution computed at the times $t_p$ corresponding
to the maximum value of $\la m^2\ra_t$.}
 \label{wf}
 \end{figure}

A complete characterization of the harmonics dynamics is obtained
by means of the harmonic probability distribution
\be
 Q(m,t) =
 \frac{\mathcal{F}(m,t)}{\sum\limits_{m\geq0}\mathcal{F}(m,t)}.
\ee
Numerical simulations given in Fig.~\ref{wf} for
$t=t_p$ show that $Q$
decreases exponentially with increasing $m$, while,
for a given $m$, the smaller the quantity $|\eta|$ is,
the larger the value of $Q$. This latter behavior is in agreement
with the fact that $A_p$ diverges when $\eta\to 0$.
Since the distribution $Q(m,t_p)$ decrease with $m$
 exponentially, we can conclude that the quantity
$\sqrt{\langle m^2 \rangle_{t_p}}$ provides a good estimate of
the number of harmonics developed up to the time $t_p$ and can
be considered as a suitable measure of the complexity
of the Wigner function at this time.

\begin{figure}
 \includegraphics[width=\columnwidth]{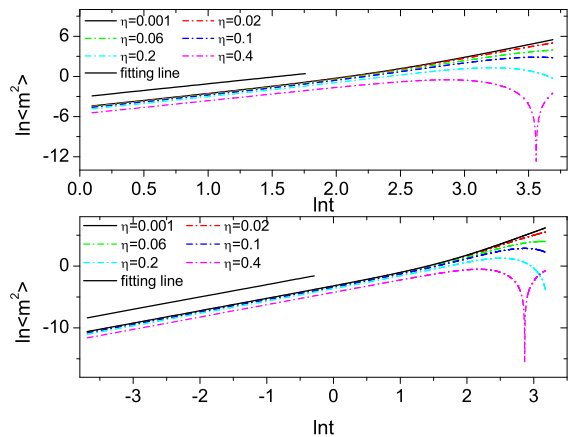}
 \caption{(Color online) Dependence of $\langle m^{2}\rangle_{t}$ on $t$
 for $\lambda_c-\lambda_0=10^{-2}$ and $\eta$ between $0.4$ and $0.001$.
 The top panel is for the normal phase and the bottom panel is
 for the superradiant phase.
 The short solid straight lines indicate the $t^2$ behavior.
 }
 \label{wgde}
 \end{figure}

Finally, we discuss our numerical simulations for the time
evolution of the number of
 harmonics of the Wigner function
 for times $t$ much shorter than the period $T$.
 Previous studies show that $\langle m^{2}\rangle_{t}$ is proportional
to $t^{2}$ in the integrable
 regime of a single-particle system~\cite{Benenti09} and grows exponentially
 in the integrable regime of a many-body system~\cite{Vinitha10}.
 In our model, we found,  both in the normal and in the superradiant phases,
that $\langle m^{2}\rangle_{t}$ increases as $t^2$ within an initial period of time
 (see Fig.~\ref{wgde}).
 For longer times (but still much smaller than $T$),
$\langle m^{2}\rangle_{t}$ increases faster than $t^2$ in both phases.
 Further discussion about these behaviors of the number of harmonics will be
 given in the following section, where the relation between this quantity
 and the LE is discussed.

\section{Lochmidt echo and fidelity for the Dicke model at QPT}
\label{sec:dickefidelity}

The divergence of the number of harmonics developed by dynamics
when approaching the critical point is a consequence
of the fact that a small difference in the controlling parameter
leads to a dramatic change of the wave function in the vicinity
of a QPT. It is therefore interesting to compare the behavior of
the number of harmonics with two other metric quantities, namely, the
quantum Loschmidt echo and the fidelity.


The so-called quantum Loschmidt echo gives a measure to
the stability of the quantum motion
under slight variation of the
Hamiltonian~\cite{peres84,nie2000,ben2004}.
It is defined by $M_L(t) = |m(t)|^2 $, where
 \be m_L(t) = \la \Psi_0|{\rm exp}(i \hat{H} t/ \hbar )
\exp(-i\hat{H}'t / \hbar)
 |\Psi_0 \ra . \label{mat} \ee
 Here
$|\Psi_0 \ra$ is the initial state,
$\hat{H}=\hat{H}'+\varepsilon \hat{V}$, and $\varepsilon$ is a small parameter.
 Extensive investigations have been performed in recent years
to understand the decaying behaviors of the LE
 \cite{jal2001,jac2001,cer2002,jac2002,cuc2002,STB03,wan2005,VH03,
 pro2002t,pro2002,wan2004,wang2005,wan2007,BC02,Gorin-rep,jacquodreport,Wang10}.
 In chaotic systems, roughly speaking, the LE has a Gaussian
decay~\cite{peres84} below a
 perturbative border and has an exponential decay
$M_L(t)\propto\exp(-\Gamma t)$ above the border.
 In the latter case, for intermediately strong perturbation~\cite{cer2002,pro2002}
 $\Gamma$ is given by the half-width of the local spectral
density of states and for relatively strong perturbation it is
 perturbation independent \cite{jal2001,STB03,wang2005}.
 In integrable systems with one degree of freedom, the LE  has a Gaussian decay,
followed after a
transient region by a power-law decay~\cite{wan2007}.
In contrast, in integrable systems with many degrees of freedom the
LE has an exponential decay~\cite{Wang10}.

 In our study of the LE,  we take $\hat{H}=\hat{H}(\lambda)$ and
$\hat{H}'=\hat{H}(\lambda_0)$, with
$\hat{H}(\lambda)$ the Hamiltonian for the Dicke model and
$\varepsilon = \lambda-\lambda_0$.
Moreover, we choose the initial state to be the ground state
$|\Phi_0(\lambda_0)\rangle$ of $\hat{H}(\lambda_{0})$,
so that the LE is in fact a survival probability.
 From Eq.(\ref{mat}) we obtain
 \be
 M_L(t)= \left| \sum_{\mu}\left|
\langle\Phi_0(\lambda_0)|\Phi_{\mu}(\lambda)\rangle\right|^2
e^{iE_{\mu}(\lambda)t}\right|^2.
 \label{ML1} \ee
 Then, substituting Eq.~(\ref{pn}) into Eq.~(\ref{ML1}), the
LE can be written as
 \be
 M_L(t)=\Big|\sum_{\mu} |C_{\mu}^{0}|^2 e^{iE_{\mu} t} \Big|^2,
 \label{mlt}
 \ee
 where $C_{\mu}^{0}$ is defined in Eq.(\ref{pn}) of Appendix A.
 Using arguments similar to those given in Sec.~\ref{sect-period-nh},
 one finds that the LE is also an oscillating function of time with
the same period
 $T=\pi/e_1({\lambda})$ as for the number of harmonics.
Moreover, $M_L(t)$ takes its minimum values, denoted by $M_p$,
 at the same times $t_p$ corresponding to the maximum values of the number of harmonics.
As shown in Ref.~\cite{QPTscaling} (see Fig. 3 therein),
the quantity $M_p$ is a function of the scaling parameter $\eta$ only,
 \be
 M_p = \frac{2\sqrt{\eta}}{1+\eta}.
\label{eq:Mp}
 \ee
The analytical derivation of this formula is here
provided in Appendix B.

 Recent investigations show that the LE may be employed to characterize QPTs
 since it has been found to have extra-fast decay in the vicinity of the critical points
 \cite{Quan06,LE-qpt,Rossini07,Peng08,Zhang09,Wang10}.
 Furthermore, as shown in Refs.~\cite{Sokolov08,Benenti09}, the number of harmonics of
 the Wigner function may be connected to the LE.
 For example, in a single-particle system,
 to the lowest order in  {$\varepsilon$}, the two quantities have the relation
 \be
 M_L(t)\simeq
1-\frac{1}{2}\varepsilon^{2}\langle{m}^{2}\rangle_{t}.
 \label{lm}
 \ee

 \begin{figure}
 \includegraphics[width=\columnwidth]{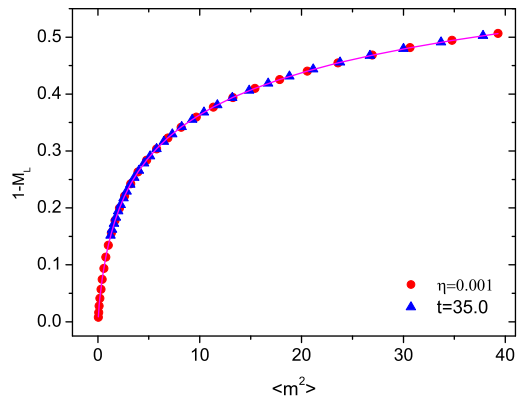}
 \caption{(Color online){ Plot of $1- M_L$ vs the number of harmonics of the Wigner function
 $\langle m^2\rangle_t$.
 Circles denote $\eta=0.001$ and $\lambda_0=\lambda_c-0.01$, with the time $t$ varying from 1 to 25.
 Triangles denote $t=35$ and $\lambda_0=\lambda_c-0.005$, with $\eta$ varying from
 0.001 to 0.3.
  The solid curve indicates the fitting function
  $1-M_L=\langle m^2\rangle_t/[3.5+\langle m^2\rangle_t+3(\langle m^2\rangle_t)^{2/3}]$.} }
 \label{wiglet}
 \end{figure}

 In order to find the relation between the number of harmonics of the Wigner
 function and the LE in the Dicke model, we have performed numerical simulations.
 We have found that there exists a
 one-to-one correspondence between the two quantities, as shown in Fig.\ref{wiglet}.
 That is, for different values of $\eta$ and $t$, the LE shows the same dependence on
 $\langle m^2\rangle_{t}$.
 Specifically, the following fitting function has been found to work well:
 \be
   M_L = \frac{a+b(\langle m^2\rangle_t)^{2/3}}
   {a+\langle m^2\rangle_t+b(\langle m^2\rangle_t)^{2/3}},
  \label{wigale}
 \ee
 with fitting parameters $a=3.5$ and $b=3.0$.
 For short times $t$ such that $[\langle m^2\rangle_t+b(\langle m^2\rangle_t)^{2/3}]
 \ll a$, Eq.~(\ref{wigale}) gives
 \be \label{mt-LE-2}
 M_L(t)\simeq 1-\frac{1}{a}\,{\langle m^2\rangle_t}.
 \ee
 Equation (\ref{mt-LE-2}) has a form similar to that in Eq.~(\ref{lm}), however, one may note a
 major difference, i.e., the quantity $a$ in Eq.~(\ref{mt-LE-2}) is a fixed (fitting) parameter,
 but not a function of the  perturbation strength $\varepsilon$.
The dependence on $\varepsilon$ in (\ref{mt-LE-2}) is contained only in
$\langle m^2 \rangle_t$ and we found numerically that
 {$\langle m^2 \rangle_t\propto \varepsilon^2$} for small  {$\varepsilon$}.

 In the Dicke model, it is known that the LE has an initial Gaussian decay \cite{Wang10,QPTscaling}.
 Then Eq.~(\ref{mt-LE-2}) implies that the number of harmonics
 $\langle m^2\rangle_{t}$ should increase as $ t^2$ within an initial period,
 in agreement with our numerical results (see Fig. \ref{wgde}) as well as
with previous results for integrable single-particle systems~\cite{Benenti09}.
 Beyond the initial Gaussian decay, after a transient region, the LE has
 been found~\cite{Wang10}
to show a $1/t$ decay for relatively long times (shorter than $T/2$) and
 for sufficiently small $\eta$.
 Consistently, Fig.\ref{wgde} shows that, in the corresponding situation,
 the number of harmonics increases faster than $t^2$.
 For $\langle m^2 \rangle_t\gg 1$,
 the scaling in Eq.~(\ref{wigale}) reduces to $M_L\simeq b \langle m^2 \rangle_t^{-1/3}$.
 Therefore, given that the LE $M_L\propto 1/t$~\cite{Wang10},
 it gives $\langle m^2 \rangle_t \propto t^3$.
 In fact, this is the reason we chose the exponent $2/3$ in the fitting in Eq.~(\ref{wigale}).
 However, we remark that, to see numerically
the dependence $\langle m^2 \rangle_t \propto t^3$, we would need
much smaller values of $\eta$ than those accessible in our
calculations.


 Finally, for the sake of completeness, we briefly review results for
 another basic metric quantity used to characterize
the occurrence of quantum phase transitions, namely, the fidelity
 \be
 L_p= |\langle \Phi_0(\lambda_0)|\Phi_0(\lambda)\rangle |,
 \label{lpe}
 \ee
 which is given by the overlap of two ground states at different
values $\lambda$ and $\lambda_0$ of the controlling parameter
\cite{gu2010,zan2006,zho2008,zan2007,coz2007,cozz2007,buo2007,zana2007}.
The dramatic change of the wave function at a QPT implies a fast
decrease of $L_p$ in the neighborhood of the critical point.
 In the Dicke model, it has been found that
 $L_p \propto |\lambda_0-\lambda_c|^{1/8}$ for a sufficiently
small $\varepsilon$~\cite{zan2006}. In Ref.~\cite{QPTscaling} we have shown
that $L_p$ only depends on the scaling parameter $\eta$, with
$ L_p = {\sqrt{2}\sqrt[8]{\eta}}/{\sqrt{\sqrt{\eta}+1}}$.

 \section{Conclusions}
\label{sec:conc}

In classical systems the growth rate of the number of
harmonics is determined by the Lyapunov exponent~\cite{Gong03}
and complexity arises from the fact that the orbits of
deterministic systems with positive Lyapunov exponent are
unpredictable, with positive algorithmic complexity~\cite{Ford,AY81}.
In the quantum realm, it was shown that the number of harmonics
can be used to measure the complexity of single-particle
systems, pure or mixed~\cite{Sokolov08}, and to detect
in the time domain the crossover from integrability to
chaos~\cite{Benenti09}.
The effectiveness of such a measure for quantum many-body dynamics
was illustrated for spin chains
in Ref.~\cite{Vinitha10}.

In this paper we have shown that the number of harmonics of the Wigner
function is a suitable quantity to characterize quantum phase transitions.
We have shown, in the case of the Dicke model, that there exists a
one-to-one correspondence between the number of harmonics
and the quantum Loschmidt echo such that both quantities can be equivalently
used to characterize the quantum phase transition.
We can conclude that the number of harmonics
emerges as an extremely broad complexity quantifier.

Finally, we point out that our analysis could be extended to
the case when several superradiant modes rather than a single one
are formed at the transition, a phenomenon in analogy
to the decay of collective excitations in highly excited
heavy nuclei~\cite{nuc1,nuc2,nuc3,nuc4,nuc5,nuc6,nuc7}.

\begin{acknowledgments}

 This work was partially supported by the Natural Science Foundation of China under Grants
 No.~11275179 and No. 10975123,  the National Key Basic Research Program of China under Grant
 No.2013CB921800, the Research Fund for the Doctoral Program of Higher Education of China,
 and the MIUR-PRIN project
``Collective quantum phenomena: From strongly correlated systems to
quantum simulators.''

\end{acknowledgments}

 \appendix

 \section{Some properties of the eigenstates of
$\hat{H}(\lambda_0)$ and $\hat{H}(\lambda)$ for the Dicke model}
 \label{eg}

 In this appendix we discuss some properties of the eigenstates of
 $\hat{H}(\lambda_0)$ and $\hat{H}(\lambda)$ for the Dicke model.
 Let us first discuss the normal phase and write the lowest-mode annihilation
 operator $\hat{c}_{1}({\lambda_0})$ in terms of the creation and annihilation operators
 for the value $\lambda$ of the controlling parameter, namely,
 \be
 \hat{c}_{1}({\lambda_{0}})= P_{1} \hat{c}_{1}(\lambda)^\dag+ P_{2}
 \hat{c}_{1}({\lambda}).
 \label{rl}
 \ee
 The coefficients $P_{1}$ and $P_{2}$ are given by~\cite{Emary03}
     \bey
     P_{1}= \frac{1}{2} \cos( {r-r_0})
     \left(\sqrt{\frac{e_{1}(\lambda_{0})}{e_{1}(\lambda)}}-
     \sqrt{\frac{e_{1}(\lambda)}{e_{1}(\lambda_{0})}}
     \right),
     \label{p11}
     \\
     P_{2}= \frac{1}{2} \cos( {r-r_0})
     \left(\sqrt{\frac{e_{1}(\lambda_{0})}{e_{1}(\lambda)}}+
     \sqrt{\frac{e_{1}(\lambda)}{e_{1}(\lambda_{0})}}
     \right),
     \label{p22}
     \eey
     where $2 {r}=\arctan\left[
4\lambda\sqrt{\omega\omega_0}/(\omega_0^2-\omega^2)\right]$.
In the limit of small $\varepsilon=\lambda-\lambda_0$,
$\cos( {r-r_0})\approx1$.

 In the close neighborhood of the critical point $\lambda_c$,
from Eq.~(\ref{ee}) one gets
 \bey
 e_{1}({\lambda}) &\simeq& \left[ \frac{8\lambda_{c}(\lambda_{c}-\lambda)
\omega\omega_{0}}{\omega_{0}^{2}+
 \omega^{2}} \right]^{1/2}.
 \label{edlt}
 \eey
 Then it is seen that $e_{1}({\lambda_{0}})/e_{1}({\lambda})\simeq
(1/\eta) ^{1/2} $, with the scaling parameter
$\eta=(\lambda-\lambda_c)/(\lambda_0-\lambda_c)$.
Using this result, we see that
 \bey
  P_{1}\simeq \frac{1}{2}
 \left(\frac{1}{\sqrt[4]{\eta}}-
 \sqrt[4]{\eta}
 \right) , \ \ \
  P_{2}\simeq \frac{1}{2}
 \left(\frac{1}{\sqrt[4]{\eta}}+
 \sqrt[4]{\eta}
 \right) ,
 \label{P12-eta} \eey
 hence $P_1$ and $P_2$ are, close to the QPT,
functions of the ratio $\eta$ only.

 We also discuss some properties of the expanding coefficients of $|\Phi_n(\lambda_{0})\rangle$
 in the basis $|\Phi_\mu(\lambda)\rangle$,
 \be|\Phi_n(\lambda_{0})\rangle= \sum_{\mu} C_{\mu}^{n}
 |\Phi_\mu(\lambda)\rangle.
 \label{pn}
 \ee
 Let us write the coefficients in the form of vectors,
 i.e., $\mathbf{C}^{n} \equiv \{C_{0}^{n}, C_{1}^{n}, C_{2}^{n},\ldots\}$.
We note that
 \bey
 \hat{c}_{1}({\lambda_{0}}) |\Phi_0(\lambda_{0})\rangle &=&0, \label{cj}\\
 \hat{c}_{1}^\dag({\lambda_{0})} |\Phi_n(\lambda_{0})\rangle
 &=& \sqrt{n+1}\ |\Phi_{n+1}(\lambda_{0})\rangle . \label{cs}
 \eey
 Substituting Eqs.~(\ref{rl}) and (\ref{pn})
(for $n=0$) into Eq.~(\ref{cj}), we obtain
 \bey
  C_{\mu}^{0} &=& \frac{-P_{1}\sqrt{\mu-1}}{P_{2}\sqrt{\mu}}\, C_{\mu-2}^{0}
 \label{c0}.
 \eey
 Then, substituting Eqs.~(\ref{rl}) and (\ref{pn}) into Eq.~(\ref{cs}),
we find
 \be
 \mathbf{C}^{n} = \frac{1}{\sqrt{n!}}\,\mathbf{C}^{0}
 (\mathbf{X})^{n},
 \label{cn}
 \ee
 where $\mathbf{X}$ is a symmetric triple-diagonal matrix
 \be
 \mathbf{X} = \left(
                \begin{array}{cccc}
                  0 & P_{2} &   &   \\
                  P_{1}& 0 &\sqrt{2}P_{2}   &   \\
                    & \sqrt{2}P_{1}& 0 &\sqrt{3}P_{2}   \\
                    &   &  \sqrt{3}P_{1} & \ddots \\
                \end{array}
              \right).
 \ee

 Next, we discuss the superradiant phase.
 Similar to Eq.~(\ref{rl}), we write
 \be
 \hat{c}_{1}({\lambda_{0}})= P_{1}' \hat{c}_{1}(\lambda)^\dag+ P_{2}'
 \hat{c}_{1}({\lambda}),
 \label{rluper}
 \ee
where
   \bey
     P^{\prime}_{1}= \frac{1}{2} \cos( {r^{\prime}-r^{\prime}_0})
     \left(\sqrt{\frac{e_{1}^{\prime}(\lambda_{0})}{e_{1}^{\prime}(\lambda)}}-
     \sqrt{\frac{e_{1}^{\prime}(\lambda)}{e_{1}^{\prime}(\lambda_{0})}}
     \right),
     \label{p1}
     \\
     P^{\prime}_{2}= \frac{1}{2} \cos( {r^{\prime}-r^{\prime}_0})
     \left( \sqrt{ \frac{e_{1}^{\prime}(\lambda_{0})}{e_{1}^{\prime}(\lambda)}}+
     \sqrt{\frac{e_{1}^{\prime}(\lambda)}{e_{1}^{\prime}(\lambda_{0})}}
     \right).
     \label{p2}
     \eey
Here $2 {r^{\prime}}=\arctan\left[
2\omega\omega_0 {\kappa^2}/(\omega_0^2- {\kappa^2}\omega^2)\right]$,
     with $\cos( {r^{\prime}-r^{\prime}_0})\approx1$
for small $\varepsilon$.
 From Eq.~(\ref{ep}), one obtains
     \bey \label{e1-super}
     e_{1}^{\prime}(\lambda) \simeq
\left[ \frac{16(\lambda-\lambda_{c})(\lambda+\lambda_{c})
(\lambda^2+\lambda^2_{c})}
     {\omega^2+\omega_0^2} \right] ^{1/2}.
     \eey
 As a result, $e_{1}^{\prime}(\lambda_{0})/e_{1}^{\prime}(\lambda)
\simeq (1/\eta) ^{1/2} $, where
 $(\lambda_{0}+\lambda_{c})(\lambda^2_{0}+\lambda^2_{c})
\simeq(\lambda+\lambda_{c})
 (\lambda^2+\lambda^2_{c})$ has been used in the neighborhood
of the critical point.
 Then $P_{1}^{\prime}$ and $P_{2}^{\prime}$ are also functions of the
ratio $\eta$ only.
 Moreover, Eq.~(\ref{cn}) also holds in the superradiant phase.

 From Eq.~(\ref{c0}) we see that the $\mu$th coordinate of
$\mathbf{C}^{n}$ is coupled to the
 $(\mu \pm 2)$th coordinates only,
 hence $C_{\mu}^{0}$ with even and odd $\mu$ are not connected.
 The continuity of the ground state of $H(\lambda)$ with respect to
variation of $\lambda$ implies that
 $C_{0}^{0}$ is considerably large at least in the case of
$\lambda \simeq \lambda_0$.
 Therefore, in the study of ground states,
only even numbers $\mu$ should contribute significantly in
Eq.~(\ref{pn}).
 Then $K_{\mu\nu}^{nm}$ in Eq.~(\ref{Q}) can be written as
 \be
 K_{\mu\nu}^{nm} =
(C^{n+m}_{\mu})^{\ast} C^{0}_{\mu} (C^{0}_{\nu})^{\ast} C^{n}_{\nu},
 \label{Qs}
 \ee
 with even numbers $\mu$ and $\nu$.

\section{Derivation of Eq.~(\ref{eq:Mp})}
 \label{mil}

 Following arguments similar to those given in Sec. \ref{sect-period-nh}
 for the times at which the number of harmonics of the Wigner function takes the
 local-maximum values,
 it is seen that the LE in the Dicke model takes its locally lowest
 values $M_p$ at the same times $t_p$.
 Therefore, we can compute $M_p$ by Eq.~(\ref{mlt}) with $t=t_p$ and obtain
 \be
 M_p=\left| \sum_{\mu} (-1)^{\mu/2}
\left| C_{\mu}^0 \right|^2 \right| ^2,
 \ee
 with $\mu$ an even integer number.
 From Eq.~(\ref{c0}) and the normalization condition, we obtain
 \be
 \left| C_{\mu}^0 \right|^2 = \frac{\mathcal{P}^{\mu/2}
\mathcal{D}_{\mu}}
 {\sum_{\mu}\mathcal{P}^{\mu/2}\mathcal{D}_{\mu}},
 \ee
 where $\mathcal{P}= (P_1/P_2)^2$ and $\mathcal{D}_{\mu} = (\mu-1)!!/(\mu)!!$.
 Noticing that
 \bey
 \sum_{\mu}\mathcal{P}^{\mu/2} \frac{(\mu-1)!!}{(\mu)!!} &=& \frac{1}{\sqrt{1-\mathcal{P}}},\\
 \sum_{\mu} (-1)^{\mu/2} \mathcal{P}^{\mu/2} \frac{(\mu-1)!!}{(\mu)!!}
 &=& \frac{1}{\sqrt{1+\mathcal{P}}},
 \eey
 we get
 \be
 M_p = \frac{1-\mathcal{P}}{1+\mathcal{P}}.
 \label{mp}
 \ee
 Then, making use of Eq.~(\ref{P12-eta}), we have
 \be
 \mathcal{P} = \left( \frac{1-\sqrt{\eta}}{1+\sqrt{\eta}}\right)^2.
 \ee
 Substituting this equation into Eq.~(\ref{mp}), we finally
obtain (\ref{eq:Mp}).


\begin{thebibliography}{99}

\bibitem{Chirikov81}
B. V. Chirikov, F. M. Izrailev, and D. L. Shepelyansky, Sov. Sci. Rev.
C {\bf 2}, 209 (1981).

\bibitem{Gu90}
Y. Gu, Phys. Lett. A {\bf 149}, 95 (1990).

\bibitem{Ford91}
J. Ford, G. Mantica, and G. H. Ristow, Physica D {\bf 50}, 493 (1991).

\bibitem{Brumer97}
A. K. Pattanayak and P. Brumer, Phys. Rev. E {\bf 56}, 5174 (1997).

\bibitem{Gong03}
J. Gong and P. Brumer, Phys. Rev. A {\bf 68}, 062103 (2003).

\bibitem{Sokolov08}
 V.~V.~Sokolov, O.~V.~Zhirov, G.~Benenti, and G.~Casati, Phys.~Rev.~E {\bf 78},
 046212 (2008).

\bibitem{Benenti09} G.~Benenti and G.~Casati, Phys.~Rev.~E {\bf 79}, 025201
 (2009).

\bibitem{SZ08}
V.~V.~Sokolov and O.~V.~Zhirov,
Europhys. Lett. {\bf 84}, 30001 (2008).

\bibitem{Vinitha10} V. Balachandran, G. Benenti, G. Casati, and
J. Gong, Phys.~Rev.~E {\bf 82}, 046216 (2010).

\bibitem{Casati12}
G. Casati, I. Guarneri, and J. Reslen, Phys. Rev. E {\bf 85}, 036208 (2012).

\bibitem{Benenti12}
G. Benenti, G. G. Carlo, and T. Prosen, Phys. Rev. E {\bf 85}, 051129 (2012).

 \bibitem{zan2006} P. Zanardi and N. Paunkovi\'{c}, Phys. Rev. E {\bf 74}, 031123 (2006).

 \bibitem{zan2007} P. Zanardi, M. Cozzini and P. Giorda, J. Stat. Mech. (2007) L02002.

 \bibitem{zana2007} P. Zanardi, H. T. Quan, X. Wang and C. P. Sun, Phys. Rev. A \textbf{75}, 032109 (2007).

\bibitem{coz2007} M. Cozzini, P. Giorda and P. Zanardi, Phys. Rev. B \textbf{75}, 014439 (2007).

 \bibitem{cozz2007} M. Cozzini, R. Ionicioiu and P. Zanardi, Phys. Rev. B \textbf{76}, 104420 (2007).

 \bibitem{buo2007} P. Buonsante and A. Vezzani, Phys. Rev. Lett. \textbf{98}, 110601 (2007).

 \bibitem{ven2007} L. C. Venuti and P. Zanardi, Phys. Rev. Lett. {\bf 99}, 095701 (2007)

 \bibitem{zho2008} H. Q. Zhou and  {J. P. Barjaktarevi\v{c}}, J. Phys. A: Math. Theor. \textbf{41}, 412001 (2008).

 \bibitem{liu2009} T. Liu, Y. Y. Zhang, Q. H. Chen and K. L. Wang, Phys. Rev. A {\bf 80}, 023810 (2009).

 \bibitem{gu2010} S. J. Gu, Int. J. Mod. Phys. B {\bf 24}, 4371 (2010)

\bibitem{rams11} M. M.~Rams and B.~Damski, Phys. Rev. Lett. {\bf 106}, 055701 (2011).

\bibitem{peres84} A. Peres, Phys. Rev. A {\bf 30}, 1610 (1984)

 \bibitem{nie2000} M. A. Nielsen and I. L. Chuang, \emph{Quantum Computation and Quantum
 Information} (Cambridge University Press, Cambridge, 2000).

 \bibitem{ben2004}
G. Benenti, G. Casati, and G. Strini,
{\it Principles of Quantum Computation and Information,
Vol. I: Basic concepts} (World Scientific, Singapore, 2004);
{\it Principles of Quantum Computation and Information, Vol. II: Basic tools and special topics}
(World Scientific, Singapore, 2007).

 \bibitem{jal2001} R. A. Jalabert and H. M. Pastawski, Phys. Rev. Lett. \textbf{86}, 2490 (2001).

 \bibitem{jac2001} Ph. Jacquod, P. G. Silvestrov, and C. W. J. Beenakker,
 Phys. Rev. E \textbf{64}, 055203(R) (2001).

 \bibitem{cer2002} N. R. Cerruti and S. Tomsovic, Phys. Rev. Lett. \textbf{88}, 054103 (2002).

 \bibitem{jac2002} Ph. Jacquod, I. Adagideli, and C. W. J. Beenakker,
 Phys. Rev. Lett. \textbf{89}, 154103 (2002).

 \bibitem{cuc2002} F. M. Cucchietti, C. H. Lewenkopf, E. R. Mucciolo,
 H. M. Pastawski, and R. O. Vallejos, Phys. Rev. E \textbf{65}, 046209 (2002).

 \bibitem{pro2002t} T. Prosen, Phys. Rev. E \textbf{65}, 036208 (2002).

 \bibitem{pro2002} T. Prosen and M. \v{Z}nidari\v{c},  {J. Phys. A: Math. Gen.} \textbf{35}, 1455 (2002).

\bibitem{BC02} G. Benenti and G. Casati, Phys. Rev.E {\bf 65}, 066205 (2002).

 \bibitem{VH03} J.~Van\'{\i}\v{c}ek and E. J.~Heller, Phys.~Rev.~E {\bf 68}, 056208 (2003).

 \bibitem{STB03} P. G.~Silvestrov, J.~Tworzyd{\l}o, and C. W. J.~Beenakker,
   Phys.~Rev.~E {\bf 67}, 025204(R) (2003).

 \bibitem{wan2004} W.-G. Wang, G. Casati and B. Li, Phys. Rev. E \textbf{69}, 025201(R)
(2004).

\bibitem{wan2005} W.-G. Wang and B. Li Phys. Rev. E \textbf{71}, 066203 (2005).

 \bibitem{wang2005} W.-G. Wang, G. Casati, B. Li, and T. Prosen, Phys. Rev. E
\textbf{71}, 037202 (2005).

 \bibitem{wan2007} W.-G. Wang, G. Casati, and B. Li, Phys. Rev. E \textbf{75}, 016201 (2007).

 \bibitem{Gorin-rep} T. Gorin, T.~Prosen, T. H.~Seligman, and M.~\v{Z}nidari\v{c}, Phys. Rep. {\bf 435}, 33 (2006).

\bibitem{jacquodreport}
Ph. Jacquod and C. Petitjean, Adv. Phys. {\bf 58}, 67 (2009).

\bibitem{Wang10}
W.-G. Wang, P. Qin, L. He, and P. Wang, Phys.~Rev.~E {\bf 81}, 016214 (2010).

\bibitem{Bargmann}
V. Bargmann, Commun. Pure Appl. Math. {\bf 14}, 187 (1961).

\bibitem{Glauber}
R. J. Glauber, Phys. Rev. {\bf 131} 2766 (1963).

\bibitem{Agarwal}
G. S. Agarwal and
E. Wolf, Phys. Rev. D {\bf 2}, 2161 (1970).

\bibitem{footnote_entropy}
The number of harmonics has been more precisely defined
in Ref.~\cite{Vinitha10} in terms of an entropic
measure. However, for the purposes of
the present paper it is sufficient to consider
the second moment $\langle \bm{m}^2 \rangle_t$,
which, as we will show, leads to a reliable estimate of
the number of harmonics for the Dicke model close to its QPT.

\bibitem{schwinger53}
J. Schwigner, Phys. Rev. {\bf 91}, 728 (1953).

\bibitem{Emary03} C.~Emary and T.~Brandes, Phys.~Rev.~E {\bf 67}, 066203 (2003).

\bibitem{QPTscaling} W.-G. Wang, P. Qin, Q. Wang, G. Benenti,
and G. Casati, Phys. Rev. E {\bf 86}, 021124 (2012).


 \bibitem{Quan06} H.~T.~Quan, Z.~Song, X.~F.~Liu, P.~Zanardi, and C.~P.~Sun,
 Phys.~Rev.~Lett.~{\bf 96}, 140604 (2006).

 \bibitem{LE-qpt} Z.~G.~Yuan, P.~Zhang, and S.~S.~Li, Phys. Rev. A {\bf 75}, 012102 (2007);
 Y. C. Li and S. S. Li, {\it ibid}.~{\bf 76}, 032117 (2007).

 \bibitem{Rossini07} D.~Rossini, T.~Calarco, V.~Giovannetti, S.~Montangero,
 and R.~Fazio, Phys. Rev. A {\bf 75}, 032333 (2007).

 \bibitem{Peng08} J.~Zhang, X.~Peng, N.~Rajendran, and D.~Suter,
 Phys. Rev. Lett. {\bf 100}, 100501 (2008).

\bibitem{Zhang09}
J.~Zhang, F.~M.~Cucchietti, C.~M.~Chandrashekar, M.~Laforest, C.~A.~Ryan,
M.~Ditty, A.~Hubbard, J.~K.~Gamble, and R.~Laflamme,
Phys. Rev. A {\bf 79}, 012305 (2009).

\bibitem{Ford}
J. Ford, Phys. Today {\bf 36}(4), 40 (1983).

\bibitem{AY81}
V. M. Alekseev and  {M. V. Yakobson}, Phys. Rep. {\bf 75}, 290 (1981).

\bibitem{nuc1}
V. V. Sokolov and V. G. Zelevinsky,
Phys. Lett. {\bf 202}, 10 (1988).

\bibitem{nuc2}
V. V. Sokolov and V. G. Zelevinsky,
Nucl. Phys. A {\bf 504}, 562 (1989).

\bibitem{nuc3}
V. V. Sokolov and V. G. Zelevinsky,
Ann. Phys. (N.Y.) {\bf 216}, 323 (1992).

\bibitem{nuc4}
F. M. Izrailev, D. Saher, and V. V. Sokolov,
Phys. Rev. E {\bf 49}, 130 (1994).

\bibitem{nuc5}
N. Lehmann, D. Saher, V. V. Sokolov, and H.-J. Sommers,
Nucl. Phys. A {\bf 582}, 223 (1995).

\bibitem{nuc6}
G. E. Mitchell, A. Richter, and H. A. Weidenm\"{u}ller,
Rev. Mod. Phys. {\bf 82}, 2845 (2010).

\bibitem{nuc7}
N. Auerbach and V. Zelevinsky,
Rep. Prog. Phys. {\bf 74}, 106301 (2011).

\end{thebibliography}
  \end{document}